\documentclass[journal]{IEEEtran}
\usepackage{cite}
\usepackage[dvipsnames]{xcolor}
\usepackage[utf8]{inputenc}
\usepackage{dirtytalk}
\usepackage{comment}
\usepackage{graphicx}
\usepackage{enumitem}
\ifCLASSINFOpdf
\else
\fi

\begin{document}

\title{An Improved Transmission Switching Algorithm for Managing Post-(N-1) Contingency Violations in Electricity Networks}

\author{Tanveer~Hussain,~\IEEEmembership{Student Member,~IEEE,}
        S M Shafiul~Alam,~\IEEEmembership{Senior Member,~IEEE,}\newline
        Siddharth~Suryanarayanan,~\IEEEmembership{Senior Member,~IEEE,}
        and~Mayank~Panwar,~\IEEEmembership{Member,~IEEE}% <-this % stops a space
\thanks{T. Hussain and S. Suryanarayanan are with the Department
of Electrical and Computer Engineering, Colorado State University, Fort Collins,
CO, USA (e-mail: tanh888@colostate.edu, sid.suryanarayanan@colostate.edu). S M S. Alam is with Idaho National Laboratory, Idaho Falls, ID, USA (e-mail: SMShafiul.Alam@inl.gov). M. Panwar is with National Renewable Energy Laboratory, Golden, CO, USA (e-mail: mayank.panwar@nrel.gov). Work supported through the INL Laboratory Directed Research \& Development (LDRD) Program under DOE Idaho Operations Office Contract DE-AC07-05ID14517.}}

\markboth{IEEE Transactions on Power Systems}%
{Hussain \MakeLowercase{\textit{et al.}}: An Improved Transmission Switching Algorithm for Managing Post-(N-1) Contingency Violations in Electricity Networks}

\maketitle

\begin{abstract}
This letter is a proof of concept for an improved transmission switching (TS) performance by moving the search space to load shed buses. Research from the past shows that changing transmission system topology changes the power flows and removes post contingency violations. Hence, TS can reduce the amount of load shed after an N-1 contingency. One of the major challenges is to find the best TS candidate in a suitable time. In this letter, the best TS candidate is determined by using a novel heuristic bi-level method based on linear sensitivity. The proposed bi-level method is easy to implement in the real world, guarantees removal of post contingency violations, and ranks the best TS candidates based on minimum load shedding possible. Moreover, the proposed method is computationally efficient since it does not involve mixed integer programming. The bi-level method is implemented by modifying the topology of transmission system after the N-1 contingency in the IEEE 39-bus test system and results show that TS with generation re-dispatch is the best solution for load shed recovery to prevent cascading failures. Moreover, the bi-level method performs even for the case when the existing methods in literature fail to completely remove post contingency violations.
\end{abstract}

\begin{IEEEkeywords}
Cascading failures, line outage distribution factor (LODF), load shed recovery, post contingency violation (PCV), transmission switching (TS)
\end{IEEEkeywords}

% \fi
\IEEEpeerreviewmaketitle

\section{Introduction}
The aim of the electrical power system is to provide continuous power to its customers. But, in case of a generator or line contingency, load shedding may be the only solution to avoid post contingency violations (PCVs). Lack of proper load shedding may lead to cascading failures. Research in the past has minimized the amount of load shedding while preventing PCVs. Refs. \cite{sinha2019optimal} and \cite{zhai2017model} provide an optimal load shedding  mechanism to mitigate cascading failures without transmission switching (TS). With the help of TS, this manuscript addresses the issue of minimizing load shedding in the aftermath of a contingency.

TS provides topology change and is beneficial for contingency analysis \cite{sadat2018reducing,7592475,7208901}. Ref. \cite{6648720} presents a DC optimal load shed recovery mechanism with TS (DCOLSR-TS), that is an NP-hard problem and faces computational issues. So, \cite{6648720} introduced a mixed integer programming heuristic algorithm (MIP-H) to reduce the complexity. But, MIP-H faces scalability issues.

Ref. \cite{7592475} developed an AC-based, computationally less complex, corrective TS based real-time contingency analysis (RTCA) tool capable of handling large-scale real power systems. Three heuristic methods (HM) proposed in \cite{7592475} are implemented on large-scale power systems and the results show that their success rate to remove PCVs completely is between 10\%-30\% \cite{7592475}. One such case is shown  in this letter where two HMs from \cite{7592475}, i.e., closest branches to contingency element (CBCE) and closest branches to violation element (CBVE), fail to remove the PCVs when implemented on the modified IEEE 39-bus system. To maintain system security, complete removal of PCVs is the ideal requirement. Which means, there is a need to improve the existing TS algorithms to mitigate cascading failures. This letter presents a bi-level algorithm which succeeds in eliminating the PCVs with optimal load shedding by relocating the search space for the solution to the load shedding buses.

\section{Proposed Algorithm}
To the best of the authors' knowledge, none of the existing algorithms in the literature uses AC optimal power flow (ACOPF) with dispatchable loads and TS to prevent cascading failures. The proposed algorithm is bi-level, as shown in Figure~1. Steps in the first level are:
\begin{itemize}
    \item start with a healthy system without any contingency.
    \item suppose a contingency happens (generator or transmission line outage).
    \item run ACOPF with dispatchable loads. The weights associated with load shedding in the objective function are higher than those associated with generators to ensure optimal load shedding.
\end{itemize}
Note that ACOPF is performed by the industry every 5-15 minutes \cite{cain2012history}. Moreover, we need to go to the second level of the algorithm if, and only if,  there is a need to do load shedding in the first level. Steps in the second level are:
\begin{itemize}
    \item \textit{find the buses where load shedding is required. We call these buses as load shedding buses (LSB)}.
    \item \textit{find the branches that are connected to LSB and are operating close to their limits. These branches are the reason for load shedding and we call them limit branches (LBr)}.
    \item from line outage distribution factor (LODF) matrix, best candidates for TS can be selected by looking at the column of corresponding LBr and choosing branches with highly negative values of LODF.
\end{itemize}
\begin{figure}[] \label{LODF}
\centering
\includegraphics[scale=0.30]{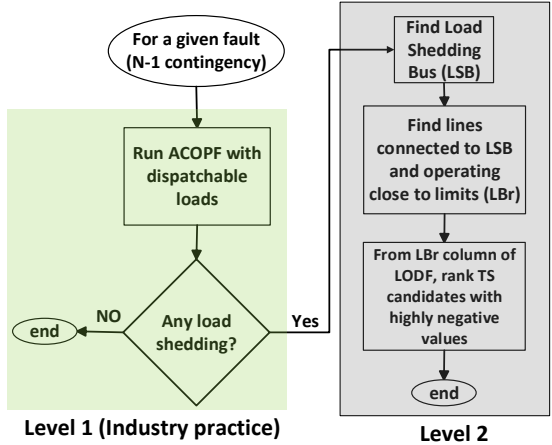}
\caption{ Proposed bi-level method for finding best TS candidates based on ACOPF and LODF }
\end{figure}

The first two steps of second level are the main contributions of this letter. LODF is an input for this algorithm and is required to be calculated only once for the given topology of the power system. Switching branch with negative value of LODF provides counter flow on LBr. The concept of finding best TS candidate by selecting highly negative value of LODF is also implemented in \cite{sadat2018reducing}. The main difference between the proposed method and the existing algorithms in \cite{7592475} is the \textit{search area}. The proposed algorithm relocates the search space to near the LSB for finding the best TS candidate whereas the algorithms proposed in \cite{7592475} search for the potential best TS candidate near a contingency element or a violation element. Moreover, the proposed method also allows for load shedding and generators re-dispatch, whereas the algorithms proposed in \cite{7592475} do not allow for either. Most importantly, our bi-level method indicates the complete removal of PCVs, whereas the algorithm in \cite{7592475} does not.
Other advantages of the proposed algorithm include:
\begin{itemize}
    \item the use of ACOPF, which is already practiced by the industry. Hence, adoption of the proposed algorithm will be easy for the industry.
    \item not involving mixed integer programming, \cite{6648720}, which makes it easier to implement.
    \item applicability to large-scale power systems.
    \item faster than the existing algorithms and ranks the best TS candidate based on minimum load shedding.
\end{itemize}

\section{Case Study} \label{case}
A popular solver of ACOPF \cite{5491276} is used to perform two case studies on the IEEE 39-bus test case system. Simulations are performed on a 2.50 GHz windows computer with 16 GB RAM. IEEE 39-bus system is N-1 complaint; hence, to show PCVs for N-1 contingency, we modified the load values as shown in Table~\ref{table1}. Note that the modified load values are uniformly distributed between [0.9, 1.1] of the nominal values. These case studies consider N-1 contingency for Branch 35 outage, which is connected between buses 21 and 22 of the IEEE 39-bus system. Branch 29, 36, and 38 are overloaded as a result of this outage. The first case study compares the results of the proposed method with \cite{7592475} and shows that the proposed method removes PCVs, whereas the existing methods in \cite{7592475} can not remove PCVs. For the second case, we compare three scenarios and show that the TS with generation re-dispatch works for load shed recovery to prevent cascading failures.
\begin{table*}[h]
\centering
\caption{Load values for IEEE 39 bus test case system}
\label{table1}
\resizebox{\textwidth}{!}{%
\begin{tabular}{|c|c|c|c|c|c|c|c|c|c|c|c|c|c|c|c|c|c|c|c|c|c|}
\hline
Bus  number     & 3   & 4   & 7     & 8   & 12   & 15  & 16  & 18  & 20  & 21  & 23    & 24    & 25  & 26  & 27  & 28  & 29     & 31  & 39   \\ \hline
Load value (MW) & 342.88 & 546.57 & 238.9 & 550.76 & 7.1977 & 311.86 & 340.83 & 153.18 & 610.2 & 298.6 & 238.5 & 334.38 & 203.02 & 126.9 & 288.3 & 221.4 & 255.64 & 9.77 & 1150.9\\ \hline
\end{tabular}%
}
\end{table*}

\begin{table}[t!]
\centering
\caption{Comparison of proposed algorithm with algorithms in \cite{7592475}}
\label{first}
\begin{tabular}{|c|c|c|c|c|}
\hline
\textbf{Algorithms}                                                & \textbf{\begin{tabular}[c]{@{}c@{}}Load\\ shedding\\ (MW)\end{tabular}} & \textbf{\begin{tabular}[c]{@{}c@{}}Post\\ contingency\\ violations\end{tabular}} & \textbf{\begin{tabular}[c]{@{}c@{}}Computation\\ time \\(sec)\end{tabular}} & \textbf{\begin{tabular}[c]{@{}c@{}}Best\\ TS\\ candidate\end{tabular}} \\ \hline
\textbf{CBCE}  & 0  & 3  & 0.32 & - \\ \hline
\textbf{CBVE}  & 0 & 3 & 0.32 & - \\ \hline
\textbf{CE} & 0 & 3  & 1.02 & - \\ \hline
\textbf{\begin{tabular}[c]{@{}c@{}}Proposed\\ method\end{tabular}} & 0 & 0  & 0.23 & 4,6,7,11,12 \\ \hline
\end{tabular}
\end{table}

\subsection{Comparison of the Proposed Method with Algorithms in \cite{7592475}}
As discussed above, PCVs can play a crucial role in cascading failures. Table~\ref{first} compares the results from using the algorithms in \cite{7592475} with the proposed bi-level algorithm. In this case study complete enumeration (CE) is also applied, i.e., removing lines, one at a time, to find the best TS candidate, along with CBCE and CBVE \cite{7592475}. In Table~\ref{first}, load shedding values for the algorithms in \cite{7592475} is zero MW because algorithms in \cite{7592475} are based on AC power flow and do not consider load shedding or generation re-dispatch. Moreover, for CBCE and CBVE, we created a list of ten branches closest to contingency/violation element. The results clearly show that there is no best TS candidate found by algorithms in \cite{7592475} that can remove PCVs. On the other hand, the proposed bi-level algorithm found the best TS candidate with complete removal of PCVs with optimal load shedding. These results also indicate that TS alone, without load shedding and generator re-dispatch may not be applicable to prevent cascading failures. Moreover, the proposed algorithm generates multiple TS candidates that work. The generation values for each TS candidate will be different. Hence, this algorithm provides flexibility to the operator to choose a TS candidate suitable for the situational needs.

\subsection{Comparison of Three Scenarios}
Here, we show the reason for including generators re-dispatch and load shedding in the proposed algorithm. Three scenarios compared in this case study are:
\begin{enumerate}[label=\Alph*)]
    \item load shedding without generators re-dispatch.
    \item load shedding with generators re-dispatch (industry practice).
    \item load shedding with generators re-dispatch and TS using CE.
\end{enumerate}
The results for the three scenarios are shown in Table~\ref{four}. scenario A shows the importance of load shedding. Load shedding, even without generator re-dispatch and TS, can remove PCVs. That means, the existing TS algorithms should consider including load shedding. Note that for all three scenarios, we considered ACOPF with load shedding. None of the three cases shows PCVs because of dispatchable loads. Comparing scenario C with CE (see Table~\ref{first}) shows the importance of generator re-dispatch and load shedding. Note that CE is based on AC power flow, which does not involve load shedding or generator re-dispatch. After any contingency in the power system, scenario B is usually practiced by the industry. From Table~\ref{four}, we observe that the amount of load shedding required by scenario B is higher than scenario C. In fact, amount of load shedding required to remove PCVs is minimum in scenario C compared to other scenarios, which makes it a potential best method.

Comparing the proposed algorithm from Table~\ref{first} with scenario C from Table~\ref{four} shows that the proposed algorithm took less time to find the same best TS candidates with no load shedding. The time taken by the proposed method is comparable with scenario B and is more than twenty times faster than scenario C. These results show that the proposed method finds the best TS candidate in almost the same time as taken by scenario B. Moreover, comparing the required amount of load shedding in scenario B with the proposed algorithm shows the load shed recovery performed by the proposed algorithm. 

\begin{table}[]
\centering
\caption{Results of three scenarios}
\label{four}
\begin{tabular}{|c|c|c|c|c|}
\hline
\textbf{Scenario} & \textbf{\begin{tabular}[c]{@{}c@{}}Load\\ shedding\\ (MW)\end{tabular}} & \textbf{\begin{tabular}[c]{@{}c@{}}Post\\ contingency\\ violations\end{tabular}} & \textbf{\begin{tabular}[c]{@{}c@{}}Computation\\ time \\(sec)\end{tabular}} & \textbf{\begin{tabular}[c]{@{}c@{}}Best\\ TS\\ candidate\end{tabular}} \\ \hline
\textbf{A} & 395.7193 & 0  & 0.23  & -  \\ \hline
\textbf{B} & 15.5963 & 0  & 0.23 & -   \\ \hline
\textbf{C} & 0  & 0 & 4.98  & 4,6,7,11,12 \\ \hline
\end{tabular}
\end{table}
\section{Conclusion}
Sometimes, load shedding might be the only solution to overcome PCVs. Research has been conducted in the past to  minimize  the amount of load shedding required to fully remove PCVs. This letter presents a novel bi-level method based on TS to remove PCVs with load shedding.  Existing TS algorithms have shown application in reducing PCVs. But, existing TS algorithms are either computationally complex, \cite{6648720}, or they fail to completely remove the PCVs and, hence can not be useful to prevent cascading failures\cite{7592475}. On the other hand, the proposed bi-level method is fast, easy to implement, and may guarantee no violations with single TS. Moreover, the proposed method is based on ACOPF with dispatchable loads and calculates multiple TS candidates based on LODF with load shedding. Also, the comparison of the proposed bi-level method with the commonly practiced method in industry, i.e., scenario B, shows that the bi-level method performs load shed recovery satisfactorily. One of the limitations of the proposed method could be the availibility of generators for re-dispatch. In the future, we plan to implement the proposed algorithm on a real world large power system to show scalability and fully address the limitations of proposed algorithm. 

\bibliographystyle{IEEEtran}
\bibliography{references}

\end{document}